\documentclass[pra,twocolumn,showpacs,superscriptaddress,preprintnumbers,amsmath,amssymb,floatfix]{revtex4}
\usepackage{bm}
\usepackage{graphicx}
\usepackage{amsbsy}
\usepackage{amsmath}
\usepackage{amsfonts}
\usepackage{amsthm}
\usepackage{color}
\usepackage{mathrsfs}
\usepackage{graphicx}
\usepackage{dcolumn}
\usepackage{bm}
\def\ket#1{|\,#1\,\rangle}
\newcommand{\bwt}{\begin{widetext}}
\newcommand{\ewt}{\end{widetext}}
\newcommand{\bea}{\begin{eqnarray}}
\newcommand{\eea}{\end{eqnarray}}

\begin{document}

\title{Photon echo quantum memory with complete use of natural inhomogeneous broadening}

\author{Sergey A. Moiseev}
\email{samoi@yandex.ru}

\affiliation{Kazan Physical-Technical Institute of the Russian Academy of Sciences,
10/7 Sibirsky Trakt, Kazan, 420029, Russia}
\affiliation{Institute for Informatics of Tatarstan Academy of Sciences, 20 Mushtary, Kazan, 420012, Russia}

\affiliation{Physical Department of Kazan (Volga Region) Federal University, Kremlevskaya 18, Kazan, 420008, Russia}

\date{\today}


\begin{abstract}

The photon echo quantum memory is based on a controlled rephasing of the atomic coherence excited by signal light field in the inhomogeneously broadened resonant line. Here, we propose a novel active mechanism of the atomic rephasing which provides a perfect retrieval of the stored light field in the photon echo quantum memory for arbitrary initial inhomogeneous broadening of the resonant line.  It is shown that the rephasing mechanism can exploit all resonant atoms which maximally increases an optical depth of the resonant transition that is one of the  critical parameters for realization of highly efficient quantum memory. We also demonstrate that the rephasing mechanism can be used for various realizations of the photon echo quantum memory that opens a wide road for its practical realization.

\end{abstract}

\pacs{03.67.-a, 42.50.Ct, 42.50.Md}


\maketitle

\section{Introduction}
\label{sec:introduction}

Quantum memory (QM) is one of the key quantum devices for practical realization of various basic protocols in quantum communication \cite{Briegel998,Kimble2008} and quantum computation  \cite{Nielsen2000,Kok2007}.  In the last decade, a considerable progress has been achieved in the optical QMs based on the atoms in cavities \cite{Cirac1997}, non-resonant Raman transitions \cite{Kuzmich2000,Julsgaard2004}, electromagnetically induced transparency \cite{Fleischhauer2000,Eisaman2005,Chaneliere2005,Novikova2007,Choi2008,Appel2008,Honda2008}, and photon-echo QM techniques
\cite{Moiseev2001,Moiseev2003,Nilsson2005,Kraus2006,Alexander2006,Riedmatten2008}.

The photon echo approach offers promising possibilities for storage of arbitrary multi-mode light fields \cite{Moiseev2007,Gisin2007,Simon2007,Nunn2008} as demonstrated recently in a storage of 64 \cite{Usmani2010} and 1090 \cite{Bonarota2010} temporal modes. Record quantum efficiencies of $69\%$ \cite{Hedges2010} and  $87\%$ \cite{Hosseini2010} have been also demonstrated for the QM of the traveling light fields in solid state and gaseous media. Moreover even higher quantum efficiency $>90\%$ is predicted for storage of 100 temporal modes in the optimal QED cavity for moderate atomic parameters \cite{Moiseev2010a}. However there are serious experimental problems in realization of the photon echo QMs with practically vital properties which are discussed in the reviews \cite{Lvovsky2009,Tittel2008,Hammerer2010,Simon2010}. Especially it is worth noting that recently developed variants of the photon echo QM use quite complicated experimental methods (see below) for realization of very delicate spectral manipulations of the inhomogeneously broadened (IB) lines which restricts the quantum efficiency of the QMs, storage time or spectral width of the signal light field. In this paper we propose a novel simple \emph{active mechanism of rephasing} (AMR-protocol) of the atomic coherence that offers new experimental possibilities for practical realization of the photon echo QM.

 In the beginning we briefly outline the experimental methods providing a temporal and spectral manipulations of the atomic coherence excited in the photon echo QM media. Then we propose a basic scheme of the AMR-protocol by using Raman type of the photon echo QM (Raman echo QM) proposed recently in \cite{Moiseev2008,Hetet2008b}, further developed  in  \cite{Nunn2008,Gouet2009} and experimentally  demonstrated in \cite{Hosseini2009,Hosseini2010}. Finally we describe how AMR-procedure can be used for original photon echo QM and discuss two perfect realizations of the photon echo QMs where AMR-protocol is protected from the negative influence of extra quantum noise. In conclusion we summarize the  advantages of AMR-protocol and outline some of its interesting applications.

 \section{Atomic rephasing in the photon echo quantum memories}
\label{sec:Rephasing procedures}

In accordance with the basic idea \cite{Moiseev2001}, the photon echo QM exploits complete absorption of the signal light pulse on the resonant IB transition providing thereby a direct pure mapping of the quantum information carried by the signal field on the excited coherence of the multi atomic ensemble.
In a free space scheme, the complete absorption of the input light pulse occurs at large optical depth for each isochromatic atomic group of the IB line which is one of the critical requirement for realization of the effective photon echo QM.
Subsequent efficient retrieval of the stored light field  is realized in the echo signal irradiated in the backward direction in comparison  with the direction of the input signal field propagation. Such scenario of the echo field generation is realized in accordance with most desirable reversible dynamics of the light field retrieval.
The retrieval is launched by inversion of the frequency detunings for each $j$-th atom $\Delta_j (t>t')=-\Delta_j (t<t')$ of the IB resonant atomic transition at some moment of time $t'$  (the procedure is called the controlled reversibility of IB (CRIB)) and provided by phasematching condition for the echo field emission.

Concrete realizations of CRIB procedure can be fulfilled by various ways, for example it occurs  automatically in the atomic gases due to opposite Doppler frequency shifts of the echo field irradiated in the backward direction to the input signal field propagation \cite{Moiseev2001}. However this scheme does not provide a long-lived QM, so it is more interesting for some quantum manipulations of the stored light field. CRIB procedure can be realized in some crystals by active inversion of local magnetic fields \cite{Moiseev2003} caused by the dipole interaction with nearest nuclei or electron spins.
Very promising CRIB procedure uses the external electric or magnetic fields for control of the solid state photon echo QM media  provided by preliminary spectral tailoring of the original IB resonant line into narrowed single pike. Here, the CRIB procedure is fulfilled  by changing a polarity of the external electric (magnetic) field gradient effecting the inversion of linear Stark (Zeeman) shifts of the atomic transition \cite{Kraus2006,Alexander2006,Tittel2008}. However, preliminary tailoring of original IB line is accompanied  by large reduction of the active atoms which  considerably reduces an effective optical depth on the atomic transition.

Reduction of the optical depth can be minimized by using so called atomic frequency comb (AFC) structure of the IB transition \cite{Riedmatten2008}, that offers promising possibilities for broad band photon echo QM \cite{Usmani2010,Bonarota2010}  demonstrated recently also for the entangled states of light \cite{Clausen2010,Saglamyurek2010}.  However, even for the ideal AFC, the optimal effective optical depth will be more than $10$-times smaller in comparison with the original optical depth of IB resonant line. Besides, some new specific experimental problems must be resolved in AFC protocol. In particular, the  retrieval time  can not be shorter of some given value determined by the AFC structure that excludes a temporal flexibility in the readout of the stored information. Perfect tailoring of the AFC structure within the IB line by using the laser hole burning technique is also a serious experimental problem in the presence of additional atomic sublevels situated closely to the active levels used in the QM.

\section{The basic equations}
\label{sec:The scheme}

Basic scheme of the light-atoms interaction for AMR-protocol is presented in Fig.\ref{Figure1}, Fig. \ref{Figure2} and Fig.\ref{Figure3}.
At time t=0 the input signal light field $\hat{A}_{1}(t,z)$
with duration $\delta t<<T_2$ ($T_2$ is a decoherent time of the Raman transition), carrier frequency $\omega_1$ and spectral width
$\delta\omega$ enters along the $+z$ direction in the medium with three-level
atoms prepared in the long-lived level $\ket{1}=\prod_{j=1}^{N} \ket{1}_{j}$. The control (writing) field with Rabi frequency $\Omega_1 $ is switched on before the entrance of the input pulse and propagates  along the wave vector $\vec{K}_{1}$ at small angle to $z$ axis with carrier frequency $\omega_1^c$.
The signal and writing fields
are in Raman resonance
$\omega_{1}-\omega_{1}^c\approx\omega_{21}$ with sufficiently large
spectral detuning $\Delta_{1}=\omega_{31}-\omega_{1}$ from the optical
transition $1\leftrightarrow 3$ so that $\Delta_{1}\gg\delta\omega, \Delta_{in}^{(31)}$ (where $\Delta_{in}^{(31)}$ is a IB for the transition $1\leftrightarrow 3$).
We assume a very weak intensity of the signal field (in particular it can be a single photon field ) so the excited population of atomic levels $2$ and $3$ can be ignored. To be concrete, below we analyze a rare-earth type of three-level scheme in the inorganic crystals where large IB $\sim 10^{8}-10^{10}$  $s^{-1}$ can be easily realized for optical transition $1 \leftrightarrow 3$ while a spectral width of the transition $1 \leftrightarrow 2$ reaches  few kilo-Hertz \cite{Tittel2008,Simon2010}. So the spectral broadening of the transition $1\leftrightarrow 2$ is neglegable in a microsecond timescale. In this case, we get the following linearized system of Haisenberg equations for the weak signal (echo) fields $\hat{A}_{12} (\tau,z)$ and for long-lived atomic coherence $\hat{R}_{12}^j$ between level $1$ and $2$:
\begin{figure}
  \includegraphics[width=0.4\textwidth,height=0.3\textwidth]{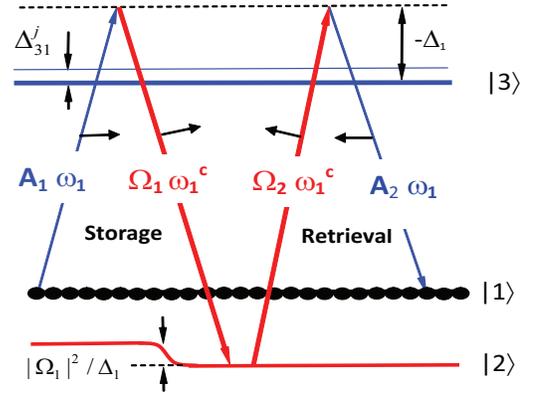}
  \caption{Energies of the atomic levels and Raman transition $1\leftrightarrow 2$ due to interaction with the probe field $A_1$ and writing field $\Omega_1$ (two left arrows) and with echo field $A_2$ and reading field $\Omega_2$ (two right arrows). Black small arrows show the wave vectors of the fields. $|\Omega_1|^2/\Delta_1$ is Stark shift of Raman transition.}
  \label{Figure1}
\end{figure}
\begin{figure}
  \includegraphics[width=0.4\textwidth,height=0.3\textwidth]{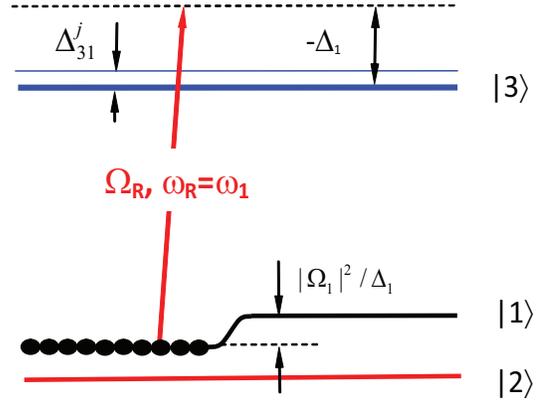}
  \caption{The rephasing pulse $\Omega_R$ is switched on adiabatically on the transition $1\leftrightarrow3$ that causes a Stark shift (dependent on $\Delta_{13}^j$ with an opposite sign in comparison with the Stark shifts induced by the control fields $\Omega_1$ and $\Omega_2$) leading to rephasing of the excited Raman coherence $\hat{R}_{12}^j$. }
  \label{Figure2}
\end{figure}
\begin{figure}
  \includegraphics[width=0.5\textwidth,height=0.25\textwidth]{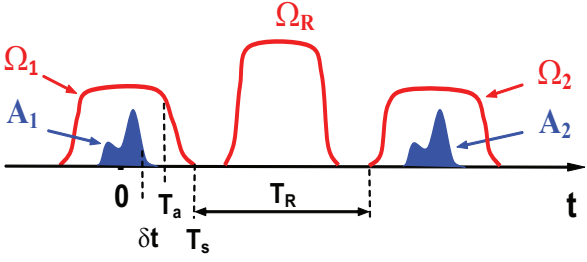}
  \caption{Basic AMR-protocol. Temporal sequence of the interaction with weak signal ($A_1$) and echo ($A_2$) fields (filled blue shapes);  writing ($\Omega_1$) and reading ($\Omega_2$) control fields are applied together with the weak light fields; $\Omega_R$ is a rephasing control laser pulse.}
  \label{Figure3}
\end{figure}
\begin{align}
\label{main A_po1}
- (-1)^{\nu}
\textstyle{\partial \over {\partial z}} \hat {A}_{\nu} (\tau_{\nu},z) =
\nonumber       \\
i\textstyle{\beta _{\nu} \over {2 }}\Big (
\chi \hat {A}_{\nu} (\tau_{\nu},z)
+ \frac{\Omega _{\nu} (t)}{g_{\nu} }
\langle\textstyle{{\hat {R}_{12}^{j} }\over {\Delta _{\nu}+\Delta_{31}^j}}\rangle
\Big ),
\end{align}

\begin{align}
\label{main R_12}
\textstyle{\partial \over {\partial \tau_{\nu}}}\hat {R}_{12}^j  = &
i\textstyle{{\Omega _{\nu}^\ast (\tau_{\nu}) g_{\nu} \hat {A}_{\nu} (\tau_{\nu},z_j )}\over{\Delta _{\nu}+\Delta_{31}^j }}
-i \delta \omega_ {\nu}(\Delta_{31}^j) \hat {R}_{12}^j,
\end{align}

\noindent
where we have used the moving system of coordinates: $t<t^{\prime}$, (${\nu}=1$), $\tau_{1}=t-z/{v_1}$, $z=z$, and for
retrieval: $t>t^{\prime}$, (${\nu}=2$), $\tau_{2}=t+Z/{v_2}$, $z=z$,
$\delta \omega_ {\nu}(\Delta_{31}^j)=\delta_{\nu}+\textstyle{{|\Omega_{\nu}(t)|^2 } f_{\nu}(\Delta_{31}^j)}$,
$\delta_{\nu}=\Delta_{21}-\textstyle{{|\Omega_{\nu}(\tau_{\nu})|^2 }\over{\Delta _{\nu}}}$,
$f_{\nu}(\Delta_{31}^j)=\textstyle{{1 }\over{\Delta _{\nu}}}-
\textstyle{{1 }\over{\Delta _{\nu}+\Delta_{31}^j }}$,
$\nu=1 (2)$ corresponds to the signal (echo) field.
For the signal and echo fields we have
$\hat{E}_{\nu}(\tau_{\nu},z)=\hat{A}_{\nu}(\tau_{\nu},z)\exp \{-i\omega _{\nu}
(t+(-1)^{\nu}n_{\nu} z/c)\}$,
$n_\nu$ is the refractive index for the signal and echo fields;
for the control fields with Rabi frequency
$\tilde{\Omega} _{\nu} (t,\vec{r}) =
 \Omega _{\nu} (\tau_{\nu} )\exp \{ - i\omega _{\nu}^c t+i\vec {K}_\nu \vec {r}\}$,
$v_{\nu} = {\partial \omega }/{\partial k}|_{\omega =
\omega _{\nu} }$ are group velocities for the signal (echo) in the absence
of interaction with atoms, and $\beta_{\nu} =2\pi (n_o S)\vert
g_{\nu}\vert ^2/{v_{\nu}}$ with atomic density $n_{o}$, photon-atom
coupling constants $g_{\nu}$, and cross section of the signal (echo)
fields $S$;
and for the atomic coherences
$\hat {P}_{12}^j (t) =
\hat {R}_{12,\nu}^j (t)\exp \{ i\varphi _{\nu} (\vec{r},z)- i(\omega
_{\nu}-\omega_{\nu}^c)(t+(-1)^{\nu}n_{\nu} z/c) \}$,
$\Delta_{21}=\omega_{21}-\omega_{1}+\omega_{1}^s$, $\varphi _{\nu} (\vec{r},z) = -((-1)^{\nu}n_{\nu}\omega _{\nu}^c z / c + \vec
{K}_{\nu} \vec {r})$,  $\vec {K}_{\nu}$ is the wave vector of the
control fields, $\langle...\rangle$ means an ensemble averaging over spectral detunings $\Delta_{31}^j$ of IB on the transition $1\leftrightarrow 3$ for atoms with spatial coordinates $z_j \approx z $:
$\langle...\rangle =\int
{d\Delta _{31,\nu}^j}$
${G(\Delta _{31,\nu}^j )...}$, $\chi=\langle\textstyle{{ 1 }\over { \Delta _{\nu}+\Delta_{31}^{j} }}\rangle$.

 In the equations (1),(2), we  have used slowly varied optical coherence $\hat R_{13}(t)$ which follows adiabatically to temporal evolution of the signal (echo) field and atomic coherence $\hat R_{12}^j (t )$ as
$\hat R_{13}^j (t) \cong \frac{g_{\nu} \hat {A}_{\nu} (t,z_j )+ \Omega
_{\nu}(t) \hat R_{12}^j (t )}{\Delta _{1} + \Delta _{31}^j  }$
 and $\hat {R}_{11}^j (t )\approx 1, \hat {R}_{22}^j (t )=\hat {R}_{33}^j (t )=0$.

\section{Storage}
\label{sec:Storage}

Similarly to the main idea of photon echo QM \cite{Moiseev2001}, we assume that  IB broadening on the Raman transition outreaches the light field width  and the resonant transition has large enough optical depth. By taking into account these spectral conditions, we launch the signal field into the medium at $\tau=0$. The probe pulse will be completely absorbed almost during the time duration of the light pulse $\tau \cong \delta t$. After the absorption we slowly switch off the control field
$\Omega_{\nu}(t > \delta t)\rightarrow 0$, so only the atomic coherence on the transition $1 \leftrightarrow 2$
will be created in the atomic system. By taking into account linear equations (1),(2) below we will analyze only the behavior of the observable values of the light field ${A}_{\nu} (\tau_{\nu},z)$   and atomic coherence  $ R_{12}^j (\tau_{\nu} )$ that is sufficient for understanding of the main properties of the analyzed QM.  By assuming that the control field amplitude
$\Omega _{\nu=1} (\tau_{\nu} )$  is constant during the interaction with probe field ${A}_{1} (\tau_{\nu},z)$, we find ${A}_{1} (\tau_{1}>\delta t,z)\cong 0$ for the optically dense media and the excited atomic coherence
${R}_{12}^j (T_{a}>\delta t)= i\textstyle{{|\Omega_{1}|^2 }\over{\Delta _{1}+\Delta_{31}^j}}
\exp\{-i \delta \omega_ {1} (\Delta_{32}^j) (T_{a}-z/{v_1}) \} \tilde {A}_{1} [\delta \omega_ {1}(\Delta_{31}^j),z]$
where

\begin{align}
\label{eq444}
\tilde{A}_{1} (\omega , z )=
\exp\{ \textstyle{\beta _{1} \over {2 }}[ i \chi - B_{1} (\omega) ] z \} \tilde{A}_{1} (\omega , 0),
\end{align}

\begin{align}
\label{eq444}
B_{1} (\omega)=
-\textstyle{{1 }\over {g_{1}}} \int {du}
\tilde G_1 (u)/[\gamma+i(\Delta_{21}-u-\omega)],
\end{align}

\noindent
where
$\tilde{A}_{1} (\omega , z )=\int_{-\infty}^{\infty} \exp\{ -i\omega \tau \}{A}_{1} (\tau , z ) {d\tau}$,
$\tilde G_1 (u)=G(\textstyle{{|\Omega_{1}|^2 }\over {u}}-\Delta_1)$, $\gamma$ is a negligibly small decay  constant of the atomic coherence.

By using Eqs. (3), (4), we find the absorption coefficient  $ \beta _{1} Re[ B_{1} (\omega)]=\textstyle{{\pi \beta _{1} }\over {g_{1}}}
G(\textstyle{{|\Omega_{1}|^2 }\over {\Delta_{21}-\omega}}-\Delta_1)$ on the frequency detuning $\omega$. The maximum absorption coefficient will be on the frequencies $\omega$ close to $\omega_o=\Delta_{21}-\textstyle{{|\Omega_{1}|^2 }\over {\Delta_{1}}}$
where the function $G(0)$ has a maximum. It is also clear that $ Im[ B_{1} (\omega_o)]\cong 0$ for small value $|\omega-\omega_o|\ll \Delta_{in}^{31}$, respectively. The value ${|\Omega_{1}|^2 }\over {\Delta_{1}}$ is a Stark shift of the IB Raman transition $1 \leftrightarrow 2$ induced by the control field  $\Omega_{1}$.

After absorption the probe field (at time $\tau_1=T_a$) we switch of the control field during $T_a<t<T_s$ for long-lived storage of the light field so
${R}_{12}^j (T_s)=
\exp\{-i \int_{T_a}^{T_s} \delta \omega_ {1} (\Delta_{31}^j, \tau_{1}) {d\tau_1} \} {R}_{12}^j (T_a)$.
We note that switching of the control field $\Omega_1$ freezes further dephasing of the atomic coherence.
Below we propose a AMR procedure for rephasing of the excited coherence $R_{12}$ by launching of one additional nonresonant control laser pulse.

\section{AMR-protocol for control of atomic coherence}
\label{sec:AMR-procedure}

The principle spectral scheme of the rephasing process is depicted in the Fig.
\ref{Figure2}. We launch a nonresonant  control light pulse $\Omega_R (\tau)$ coupling only the atomic levels $1$ and $3$. Carrier frequency of the rephasing pulse coincides with carrier frequency of the signal field. It is well-known that the selective interaction of the control field with the transition $1\rightarrow3$ can be experimentally realized by exploiting the properties of allowed and forbidden atomic transitions or frequency vicinity between the carrier frequency and the atomic transition (see also below). Here, the evolution of atomic coherence $R_{12}^j$ is determined by the following equation.

\begin{align}
\label{main R_12}
\textstyle{\partial \over {\partial \tau_{\nu}}}{R}_{12}^j  = &
-i \delta \omega_ {R}(\Delta_{31}^j,\tau)  {R}_{12}^j,
\end{align}

\noindent
where the frequency detuning

\begin{align}
\label{main delta_R}
\delta \omega_ {R}(\Delta_{31}^j,\tau)=\delta_{R} (\tau)-\textstyle{{|\Omega_{R}(\tau)|^2 } f_{1}(\Delta_{31}^j)},
\end{align}

\noindent
where
$\delta_{R}(\tau)=\Delta_{21}+\textstyle{{|\Omega_{R}(\tau)|^2 }\over{\Delta _{1}}}$.
We note that the frequency shift $\delta \omega_ {R}(\Delta_{31}^j,\tau)$ gets  an opposite frequency dependence on the atomic detuning $\Delta_{31}^j$ in comparison with the frequency shift in Eq. (2). This is a result that the
rephasing nonresonant field couples the atomic states $1 \leftrightarrow 3$ but not the states $2 \leftrightarrow 3$.
Here, we have also assumed a slowly (adiabatically) varying amplitude of the control field $\Omega_{R}(\tau)$ that excludes any real atomic transition $1\leftrightarrow3$. However, below we also discuss an additional method which have to be applied for complete elimination of negative influence caused by the spontaneous induced transition  $1\rightarrow 2$ during the rephasing procedure.

We apply the rephasing field $\Omega_{R}(\tau)$ only for finite temporal duration $T_R$. The switching of the rephasing pulse results to the following atomic coherence

\begin{align}
\label{main R_12}
{R}_{12}^j (T_s+T_R)=
\exp\{-i \int\limits_{T_s}^{T_s+T_R} \delta \omega_ {R} (\Delta_{31}^j, \tau) {d\tau} \} {R}_{12}^j (T_s)
\nonumber \\
=i\textstyle{{|\Omega_{1}|^2 }\over{\Delta _{1}+\Delta_{31}^j}}
\exp\{-i \theta +i f_{1} (\Delta_{31}^j)P(S,R) \}
\tilde {A}_{1} [\delta \omega_ {1}(\Delta_{31}^j),z],
\end{align}

\noindent
where
$\theta=\int_{-\infty}^{T_s} \delta_ 1  {d\tau} +\int_{T_s}^{T_s+T_R} \delta_R {d\tau}$ is a constant phase shift,
the factor $P(S,R)=\int_{T_s}^{T_s+T_R} |\Omega_{R} (\tau)|^2 {d\tau}
- \int_{-\infty}^{T_s} |\Omega_{1} (\tau_1)|^2 {d\tau_1}$
determines conditions of the atomic rephasing. For some fixed temporal duration $T_R'$, the factor $P(S,R)=0$ that means a complete recovering of the atomic coherence ${R}_{12}$. Below we use larger temporal duration $T_R>T_R'$ where the rephased coherence is realized again but with opposite atomic phase shifts for each isochromatic group. For simplicity we use  equaled magnitudes of the control fields $\Omega_1 = \Omega_R$ with adiabatic switching of the rephasing pulse at time $t=T_s+T_R$ ($\Omega_{R} (\tau >T_s+T_R)=0$). By assuming a large enough temporal duration $T_R$ (for example $T_R=2 T_s$ or larger) we have prepared the atomic system ($R_{12}$) for readout of the stored signal light field.

\section{Echo signal irradiation}
\label{sec:Echo irradiation}

Here we launch the readout control pulse $\Omega_{2} (\tau_2)$ at $\tau_2>T_s+T_R$ in the almost opposite direction in comparison with the first writing control pulse in order to provide phasematching condition and propagation of the echo field in the backward direction to the signal light pulse (see also the details in \cite{Moiseev2008}). In this case we prepare the initial atomic state on the second ground level and exploit larger wave vectors of the writing and reading control laser fields that leads us the following initial atomic coherence
${R}_{12,in}^j (T_s+T_R) ={R}_{12}^j (T_s+T_R)\exp\{i \delta k(\Delta_{31}^j) z\}$, where the appropriate value of $\delta k$ provides the phasemismatch condition (see below) due to using an difference of energies between level 1 and 2. In order to satisfy a temporally reversible behavior we exploit the same amplitude of the reading control field
$\Omega_{2} =\Omega_{1}$ during the echo signal emission and the  same frequency detuning $\Delta_2=\Delta_1$.

Evolution of the light field dynamics is determined by the equations (1) and (2) with index $\nu=2$. Initially, the launched reading pulse will only recover the macroscopic atomic coherence ${R}_{12}$ during temporal interval $T_s-T_a$ so the complete rephasing of the atomic coherence will occur later at $t\cong T_R+T_s=3 T_s$. By taking into account the initial state in Eq. (7), we find the following equation for the Fourier component of the echo field $\tilde {A}_{2} (\omega,z)$

\begin{align}
\label{main A_echo}
- \textstyle{\partial \over {\partial z}} \tilde {A}_{2} (\omega,z) =
\textstyle{\beta _{\nu} \over {2 }} \Big ( i \chi
-B_1 (\omega) \Big )\tilde {A}_{2} (\omega,z)
\nonumber       \\
-\exp\{i(\omega  T_R  -\theta)\}
\}\frac{\pi\beta}{g_1}\int \frac{{d\Delta} e ^{i\delta k (\Delta) z }G(\Delta) |\Omega_1|^2}{(\Delta_1+\Delta)^2}
\nonumber       \\
\exp\{+i P (S,R) f_1(\Delta)+\textstyle{\beta _{\nu} \over {2 }}\Big ( i  \chi-B_1[\delta\omega_1(\Delta)] \Big )z \}
\nonumber       \\
\delta(\omega-\delta\omega_1(\Delta))
\tilde {A}_{1} (\delta\omega_1(\Delta),0),
\end{align}

\noindent
where $\theta$ is some constant phase shift.

By integrating (8)  over the delta-function $\delta(\omega-\delta\omega_1(\Delta))$ with substitution $u=\textstyle{{|\Omega_1|^2} \over {\Delta+\Delta_1}}$, we find the following solution

\begin{align}
\label{A_echo}
\tilde {A}_{2} (\omega,z=0)=
-\exp\{-i\theta+i \omega ( T_R +T_s)\}
\nonumber       \\
\textstyle{{\pi  }\over {g_{1}}}
\frac{G(\textstyle{{|\Omega_{1}|^2 }\over {\Delta_{21}-\omega}}-\Delta_1)}
{\Big( B_1 (\omega)-i[\chi+\delta k(\omega)/\beta _{1}] \Big)}
\tilde {A}_{1} (\omega,0),
\end{align}

\noindent
 where we have taken into account
$T_s=P (S,R)/|\Omega_1|^2$ and large optical depth of the Raman transition
 $Re[ B_1 (\omega)] \beta L >>1$.

The function
$G(\textstyle{{|\Omega_{1}|^2 }\over {\Delta_{21}-\omega}}-\Delta_1)$ reaches a maximum while
$Im [B_1(\omega)]\cong (\omega-\omega') B_1(\omega ')_{\omega'} ' $ close to the frequency detuning $\omega '=\Delta_{21}-|\Omega_1|^2/{\Delta_1}$ (center of the input pulse spectrum).
 Therefore as it is seen in Eq.(9), we can satisfy the phase matching condition  by using of the relation  $\delta k(\omega) \cong  -\beta_1 \chi+ \delta\kappa '_{\omega '} (\omega-\omega ')$.
So the denominator in Eq.(9) can be simplified for narrow spectral width of the input light field as follows
\begin{align}
\label{denominator}
{\Big( B_1 (\omega)-i[\chi+\delta k(\omega)/\beta _{1}] \Big)}\cong
\nonumber       \\
\textstyle{{\pi  }\over {g_{1}}} G(\textstyle{{|\Omega_{1}|^2 }\over {\Delta_{21}-\omega}}-\Delta_1)
+i (\omega-\omega ') [B_1(\omega ')_{\omega'} '- \delta\kappa '_{\omega '}]\cong
\nonumber       \\
\textstyle{{\pi  }\over {g_{1}}}G(\textstyle{{|\Omega_{1}|^2 }\over {\Delta_{21}-\omega}}-\Delta_1)
\exp\{i (\omega-\omega ') \delta \tau\},
\end{align}
\noindent
where $\delta \tau\cong \textstyle{{g_{1}  }\over {\pi}}
(B_1(\omega ')_{\omega'} '- \delta\kappa '_{\omega '}/\beta _{1})/{G(0)}$. Finally after the Fourier transformation we find the echo field

\begin{align}
\label{A_echo_t}
 {A}_{2} (\tau,z=0)=
 \nonumber       \\
-\exp\{i (\omega '\delta \tau-\theta)\}
{A}_{1} (\tau-T_R-T_s+\delta\tau,z=0).
\end{align}

As seen in Eq.(10), the echo field completely reproduces the input signal field similar to AFC protocol while we remind that usual scenario of the photon echo QM  \cite{Moiseev2001} leads to the temporally reversed shape to the signal field. The original temporal shape of the echo field in AMR-protocol is caused by the same temporal behavior of the atomic coherence on each spectral component of the IB line. Here, we have to note that absence of the temporal reversibility in light-atoms dynamics can lead to unreversible behavior due to spectral dispersion in echo field emission (see denominator in Eq.(9)). However the weak dispersion leads only to an additional time delay $-\delta\tau$ and phase shift $\omega '\delta \tau$ which is possible for narrow enough spectral width of the signal field.

The described scheme of QM needs additional analysis and some improvement since  the rephasing laser pulse induces spontaneous Raman transitions $|1>\rightarrow |2>$ so a direct use of the schemes depicted in Figs.1-3  leads to extra quantum noises in the irradiated echo field.
Below  we describe the procedure providing complete elimination of any drawbacks caused by the spontaneous transitions.

\begin{figure}
  \includegraphics[width=0.4\textwidth,height=0.4\textwidth]{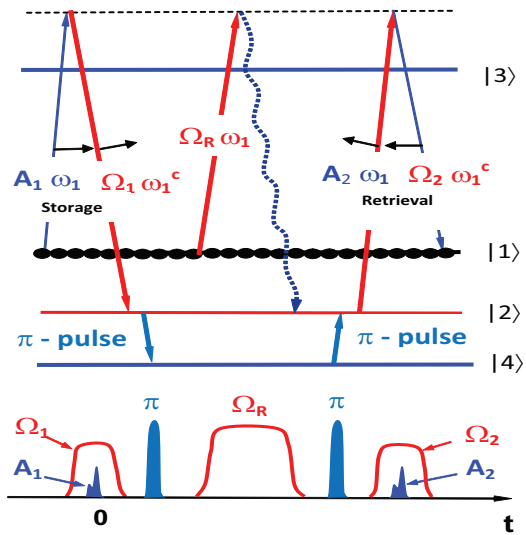}
  \caption{Atomic transitions in four level systems where first two arrows indicate the input $A_1$ and writing $\Omega_2$ fields, then $\pi$- pulse transfer the atoms from level 2 to the level 4; laser pulse $\Omega_R$ rephases the atomic coherence $R_{14}$ and leads to spontaneous transitions of atoms on level 2 (dotted waved arrow directed to level 2); the additional atomic population $\delta\rho$ of level $2$ is transferred to the level $4$ by second $\pi$ pulse; last two arrows indicate the reading $\Omega_2$ and echo $A_2$ fields.}
  \label{Figure4}
\end{figure}

Let us consider four level realization of the described QM protocol depicted in Fig.\ref{Figure4}  where an additional (buffer) level 4 could be some hyperfine sublevel similar to other ground levels 1 and 2.
Here, we assume that the transition $|1>\leftrightarrow |4>$ is forbidden for the Raman transitions realized during the storage and echo field retrieval. Before rephasing of the excited atomic coherence $R_{12}$ by the  laser pulse $\Omega_R$, we transfer the coherence $R_{12}$ to the long-lived coherence  $R_{14}$ by resonant $\pi$-pulse on the transition $|2>\leftrightarrow |4>$. Then we apply the laser pulse $\Omega_R$ which rephases now the coherence $R_{14}$  that follows the equations similar to Eqs.(5),(6). Rephasing of coherence $R_{14}$ is accompanied by some additional  population $\delta\rho_2$ of level $2$ due to the spontaneous Raman transitions caused by the rephasing pulse $\Omega_R$ (see Fig.\ref{Figure4}). After that we can apply second $\pi$-pulse on the transition $|2>\leftrightarrow |4>$ for transfer of the rephased coherence $R_{14}$ to the coherence $R_{12}$ and remove the atomic population $\delta\rho_2$ to the level $4$.  Thereby the coherence $R_{12}$ is prepared for retrieval of the stored information in the echo field $A_2$ without any quantum noise since all the atoms excited by the  spontaneous Raman transitions will stay on the level 4.

Finally we note that the described procedure of AMR can be applied for original scheme of the photon echo QM \cite{Moiseev2001,Moiseev2003} where the input signal pulse is absorbed on the optical transition $|1>\leftrightarrow |3>$. In this case we have to use large enough spectral detuning $\Delta_1 \gg |\Delta_{13}^j|$ for the rephasing pulse so that the frequency detunings during the rephasing stage will be given by ${\Omega_R^2}/(\Delta_1 +\Delta_{13}^j)\cong {\Omega_R^2}/{\Delta_1}-\Delta_{13}^j|\Omega_R/{\Delta_1}|^2$. Here, we get the same spectral shape of IB as it takes place for the absorption of the signal light field. Moreover by taking into account that factor $f_R=|\Omega_R/{\Delta_1}|^2$ can be close to unity, we can rephase the excited atomic coherence $R_{13}$ (after transfer to the long lived coherence $R_{12}$)  within the same temporal scale and eliminate thereby the negative drawbacks caused by the spontaneous transitions on the level $2$ as it is described in previous section.

\section{Conclusion}
\label{sec:Conclusion}

We have described a novel simple scheme of the photon echo QM where the rephasing of the  atomic coherence (AMR procedure) is realized by using of additional nonresonant interaction with  control laser field.  We have also demonstrated that the proposed atomic rephasing can be realized without negative influence of the quantum noises by using an additional buffer level $4$ for the atoms excited by the spontaneous Raman transitions during the rephasing stage.
We have shown that AMR procedure can be used for Raman echo QM and for usual photon echo QM. In the last case we use larger spectral detuning $|\Delta_{31}^j|<<\Delta_{1}$ where the Raman transition will get a spectral IB shape which is differed only by the factor $f_R=|\Omega_R|^2/\Delta_1|^2$ from the original shape of IB line. Therefore the time of echo field irradiation will be scaled only by the factor $f_R$.

The proposed AMR procedure provides a possibility of photon echo QMs for atomic systems with arbitrary inhomogeneous  broadenings that offers now new practical perspectives for realization of the efficient optical quantum memories and repeaters.
We believe that the proposed scheme of photon echo QMs will be interesting for quantum manipulations of the light fields, in particular for the purposes of quantum compression \cite{Hosseini2009,Moiseev2010} and frequency conversion \cite{Moiseev2008}.  We also anticipate considerable advantages of AMR procedure for the Raman echo QM on surface plasmon polariton fields which is very promising for nanoscale storage of the light fields \cite{Moiseev2010b}.

\section{Acknowledgement}
\label{sec:Ack}

Financial support by the Russian Fund for Basic Research grant \#
10-02-01348-a and government contract of
RosNauka 02.740.11.01.03 is gratefully acknowledged.


\section{References}

\begin{references}

\bibitem{Briegel998} H.-J. Briegel, W. D\"{u}r, J.I. Cirac, P. Zoller, \prl \textbf{81}, 5932 (1998).
\bibitem{Kimble2008} H.J. Kimble, Nature \textbf{453}, 1023 (2008).
\bibitem{Nielsen2000}M.A. Nielsen and I.L. Chuang 2000, Quantum Computation and Quantum Information (Cambridge Univ. Press).
\bibitem{Kok2007} P. Kok, W.J. Munro, K. Nemoto, T.C. Ralph, J.P.Dowling, G.J. Milburn,
\rmp \textbf{79}, 135 (2007).
\bibitem{Cirac1997} J.I. Cirac, \textit{et al.}, \prl \textbf{78}, 3221 (1997).
\bibitem{Kuzmich2000} A. Kuzmich, and E.S. Polzik, \prl \textbf{85}, 5639 (2000).
\bibitem{Julsgaard2004} B. Julsgaard, \textit{et. al.}, Nature \textbf{432}, 482 (2004).
\bibitem{Fleischhauer2000} M. Fleischhauer, and M.D. Lukin, \prl \textbf{84}, 5094 (2000).
\bibitem{Eisaman2005} M.D. Eisaman, \textit{et. al.}, Nature \textbf{438}, 837 (2005).
\bibitem{Chaneliere2005}T. Chaneliere, \textit{et. al.}, Nature \textbf{438}, 833 (2005).
\bibitem{Choi2008} K.S. Choi,  \textit{et. al.}, Nature \textbf{452}, 67 (2008).
\bibitem{Novikova2007} I. Novikova, \textit{et. al.},  \prl {\bf 98}, 243602 (2007).
\bibitem{Appel2008} J. Appel, \textit{et. al.}, \prl {\bf 100}, 093602 (2008).
\bibitem{Honda2008} K. Honda, \textit{et. al.}, \prl \textbf{100}, 093601 (2008).
\bibitem{Moiseev2001} S.A. Moiseev, and S. Kr\"{o}ll, \prl \textbf{87}, 173601 (2001);
 S.A. Moiseev, and B.S.Ham, Phys.Rev.A. \textbf{70}, 063809, (2004).
\bibitem{Moiseev2003}S.A. Moiseev, V.F. Tarasov, and B.S. Ham, J.Opt.B: Quantum Semiclass. Opt. 5, S497 (2003).
\bibitem{Nilsson2005} M. Nilsson, and S. Kr\"{o}ll, Opt. Commun. \textbf{247}, 393 (2005).
\bibitem{Kraus2006} B. Kraus, \textit{et. al.}, \pra \textbf{73}, 020302 (2006).
\bibitem{Alexander2006} A.L. Alexander \textit{et al.}, \prl \textbf{96}, 043602 (2006); G. H\'{e}tet,  \textit{et. al.}, \prl {\bf 100}, 023601 (2008).
\bibitem{Riedmatten2008} H. De Riedmatten, \textit{et al.}, Nature \textbf{456}, 773 (2008); M. Afzelius \textit{et. al.}, \pra \textbf{79}, 052329 (2009).
\bibitem{Moiseev2007} S.A. Moiseev, J. of Phys. B: Atom., Mol. and Opt. Phys. \textbf{40}, 3877 (2007).
\bibitem{Gisin2007} N. Gisin, S.A. Moiseev, and C. Simon, \pra \textbf{76}, 014302 (2007).
\bibitem{Simon2007} C. Simon, \textit{et. al.}, \prl \textbf{98}, 190503 (2007).
\bibitem{Nunn2008} J. Nunn, et.al., \prl \textbf{101}, 260502 (2008).
\bibitem{Usmani2010} I. Usmani, M. Afzelius, H. de Riedmatten,  and N. Gisin, Nat. Commun, \textbf{1}, 1 (2010).
\bibitem{Bonarota2010} M. Bonarota, J.-L. Le Gouet, T. Chaneliere, arXiv: 1009.2317v1 [quant-ph].
\bibitem{Hedges2010} M.P. Hedges, \textit{et. al.}, Nature,  \textbf{465}, 1052 (2010).
\bibitem{Hosseini2010}M. Hosseini, B.M. Sparkes, P.K. Lam and B.C. Buchler, arXiv: 1009.0567v1 [quant-ph].
\bibitem{Moiseev2010a} S.A. Moiseev, S.N. Andrianov, and F.F. Gubaidullin. \pra \textbf{82}, 022311 (2010).
\bibitem{Lvovsky2009} A.I. Lvovsky, B.C. Sanders, W. Tittel, Nature Photon. \textbf{3}, 706 (2009).
\bibitem{Tittel2008} W. Tittel,  M.Afzelius, T. Chaneliere, R.L.Cone, S.Kroll, S.A.Moiseev, and M.Sellars,
Laser \& Phot. Rev, \textbf{4}, 244 (2010).
\bibitem{Hammerer2010} K. Hammerer, A.S. S\"{o}rensen and E.S. Polzik, \rmp \textbf{82}, 1041 (2010).
\bibitem{Simon2010} C. Simon, \textit{et. al.},
Eur. Phys. J. D \textbf{58}, 1 (2010).
\bibitem{Clausen2010} C. Clausen, \textit{et. al.}, arXiv:1009.0489v2 [quant-ph].
\bibitem{Saglamyurek2010} E. Saglamyurek,\textit{et. al.}, arXiv:1004.4691 [quant-ph].
\bibitem{Moiseev2008} S.A. Moiseev, and W. Tittel,  arXiv:0812.1730v2 [quant-ph].
\bibitem{Hetet2008b} G. H\'{e}tet, \textit{et. al.}, Opt. Lett. \textbf{33}, 2323 (2008).
\bibitem{Gouet2009}J.L. Le Gou\"{e}t, and P.R. Berman, \pra \textbf{80}, 012320
(2009).
\bibitem{Hosseini2009} M. Hosseini, et al., Nature \textbf{461}, 241 (2009).
\bibitem{Moiseev2010} S.A. Moiseev, W. Tittel, \pra \textbf{82}, 012309 (2010).
\bibitem{Moiseev2010b} S.A. Moiseev, and E.S. Moiseev, Proceedings of the NATO Workshop: "Quantum Cryptography and Computing" (R. Horodecki, S.Ya. Kilin, J. Kowalik (eds.), IOS Press, 2010, p.212).
\end{references}
\end{document}